# Towards the fabrication of a morphing plasma source for biomedical applications

Carles Corbella[1,a], Sabine Portal[1], Li Lin[1], Michael Keidar[1]

[1] Department of Mechanical & Aerospace Engineering, George Washington University, 800 22nd Street, Northwest, Washington, DC 20052, United States of America

A new design of plasma source that merges the main features of capacitive dielectric barrier discharge (DBD) and cold atmospheric plasma jet (CAPJ) is discussed. The DBD system consists of a flexible, porous matrix consisting of silica aerogel, which is comprised between two planar-parallel electrodes. The supply of helium flow submitted to a sinusoidal voltage of 5-9 kV in amplitude and ≈15 kHz in frequency provides a 2D-distribution of plasma jets that propagate around 1 cm beyond the active DBD region. The plasma multi-jet system is aimed at surface treatment of 3D objects and large areas. CAPJ performance as a hypothetical morphing source in flat and bent configurations is discussed. Optical emission spectroscopy (OES) diagnostics has provided the composition of the CAPJ through the aerogel layer. This novel source is promising for situations requiring thorough adaptation of plasma parameters to delicate samples, as for example wound healing and treatment of surgical margins in plasma-based cancer surgery.

---

[a] Author to whom correspondence should be addressed. Electronic mail: ccorberoc@gwu.edu.

An important challenge in plasma medicine applications consists of developing new plasma sources adaptable to soft surfaces that show non-trivial topologies, like asperities, sharp edges, cavities or complex accessibility.[1,2] Two main approaches in plasma sources for biomedicine have been considered so far: cold atmospheric plasma jet source with all integrated components (CAPJ concept) and floating-electrode DBD plasma source (FE-DBD concept).[3-6] CAPJ devices provide a well defined plasma effluent and are very convenient for local plasma treatment, as they are restricted to small areas (~1 mm$^2$).[4] Such a feature becomes a downside when it is necessary to irradiate large areas homogeneously. Therefore, a number of CAPJ arrays and other sophisticated architectures have been developed to increase the surface area and homogeneity of the interacting beam, especially aiming at decontamination, sterilization and disinfection applications. The objective of the reported prototypes has consisted of fabricating either arrays of plasma jets,[7,11-14] or flexible DBD sources,[8-10] always conformal with the treated substrate. However, scanning routines are usually required to improve plasma uniformity over extended surfaces. It is thereby convenient to design a thin, flexible and adaptable plasma source to treat large area surfaces, which allows to joining the advantages from multi-jets and flexible DBDs.

In the present article, we report on the characterization of a flexible, extended CAPJ source able to mimic surfaces with non-flat topography for plasma treatments in open air. Aerogel meets all these properties. Such a device concept is new and necessary for the above reasons, and it constitutes the first step towards the design of a multi-jet CAP device with morphing capabilities. A flexible and porous silica layer from Kudosale (thicknesses from 2 to 4 mm with 0.5 mm precision), sandwiched between two planar-parallel spacers with electrodes, is connected to a He gas supply system in order to provide parallel particle beams with sufficient flow rate and energy (Fig. 1). Silica aerogel has been selected as dielectric barrier material because of its flexibility, high electrical resistivity and property of refractory material. The aerogel CAPJ matrix exhibits a mesoporous structure reinforced with microscopic fiberglass, as verified via scanning electron microscopy (SEM). Its chemical composition was assessed by energy-dispersive X-ray spectroscopy (EDS): the stoichiometry of Si and O was 1:2, with a small concentration of impurities (10 at.% C). The aerogel porosity is estimated to be around 90% via microbalance measurements and comparing the measured density with the density of ceramic silica. However, this material shows a reduced gas permeability due to the very small pore size, in the nanometer range.[15] Indeed, ceramic barriers with at least sub-millimeter-sized pores have been required for an adequate gas flow permeation in CAPJ formation.[16] The increase in permeability, so that the production of multiple jets is feasible, is achieved in the present study by puncturing a pattern of pin-hole orifices (0.5 mm in diameter) through the aerogel barrier. A DBD plasma can be then generated at the level of the pin-holes and projected as a CAP multi-jet along the He flow stream.

The details of the electrical circuit powering the multi-jet are described elsewhere.[17] Briefly, the constant voltage from a DC power supply is modulated by means of the AC signal from a function generator. The frequency range between 14 and 15 kHz yielded stable CAPJs. The resulting waveform is transformed into a high-voltage sinusoidal signal (5-9 kV in amplitude, 25-30 W in supplied power) and is applied to the CAPJ electrodes. A full description of the current-voltage characteristics will be addressed in a separate publication. The sample holder containing the aerogel matrix consists of two planar polymer spacers fabricated via 3D printing (Fig. 1a). The inlet gas tube was directly connected into a circular aperture of 8 mm in diameter at the side of the biased electrode (Fig. 1b). The outlet nozzle, which has 10 mm in diameter, is surrounded by a circular electrode connected to ground (Fig. 1c). Helium (purity: 99.995%) was provided at a total flow rate of between 1 and 2 slm. At flow rates higher than 2 slm, the performance of the plasma jet was degraded (instable jet) because the gas flow transitioned from a laminar pattern to a turbulent regime.[18] On the other hand, lower flow rates tend to uniformize the flow distribution thanks to an increase in the gas residence time at the dielectric barrier level.

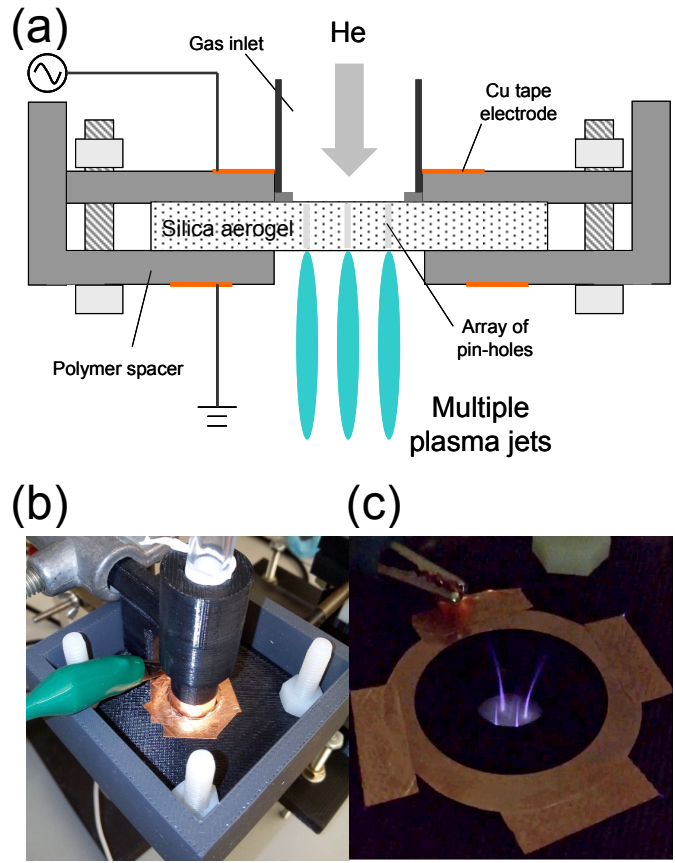

**FIG. 1. (a)** Cross-section sketch of the flexible plasma source. In the shown downstream mode, upper electrode is biased with AC high voltage and lower electrode is grounded. **(b)** Rear view (biased electrode) with the gas input, and **(c)** front view (grounded electrode) of the setup with four CAPJs ignited.

Fig. 1c shows the CAP multi-jet system with four active jets in upstream mode. The length of the jet plume is of the order of 1 cm. The observed divergence of the jet columns is basically attributed to the gas flow expansion from the nozzle. Electrostatic interaction between the charged streamers constitutes another source of jet divergence.[19] Further examples of multi-jet arrangements have been tested. In particular, we proved that the source is fully operative with five jets under any orientation of the setup: upstream, downstream and sideways. Note that the jet instabilities and secondary discharges at the electrodes were observed when working at frequencies lower than 14 kHz. The CAPJ nozzle remained at room temperature during operation, as measured with a contact-less infrared thermometer.

The optical characterization of the jet plume was performed by OES measurements, which provided the optical emission spectrum displayed in Fig. 2. The UV-visible spectrum of the air/He plasma was obtained by means of a StellarNet spectrometer operating at a spectral range 191.0-851.5 nm with a spectral resolution of 0.5 nm. For this characterization, a single CAPJ was ignited through the aerogel using a 2 mm-hole in order to obtain a brighter light source. The spectrum was measured at approximately axial distance 5 mm (middle plume length) and radial distance 5 mm from the nozzle with an integration time of 2 s. The emission lines have been attributed to the following transitions: 308.9 nm OH ($A^2\Sigma^+ \rightarrow X^2\Pi$, $v'=v''=0$), 337.13 nm $N_2$ ($C^3\Pi_u \rightarrow B^3\Pi_g$, $v'=v''=0$), 357.69 nm $N_2$ ($C^3\Pi_u \rightarrow B^3\Pi_g$, $v'=0$, $v''=1$), and the interval 370-390 nm of other $N_2$ ($C^3\Pi_u \rightarrow B^3\Pi_g$) combined with $N_2^+$ ($B^2\Sigma_u^+ \rightarrow X^2\Sigma_g^+$) transitions. The typical CAPJ contributions at 706.519 nm He ($3^3S \rightarrow 2^3P$) and ≈777 nm O ($3^5P \rightarrow 3^5S$) are under the detection threshold.[20,21] A spatially-resolved optical diagnostics along the jet plume, as well as the determination of plasma parameters (temperature and density), are planned for future work.

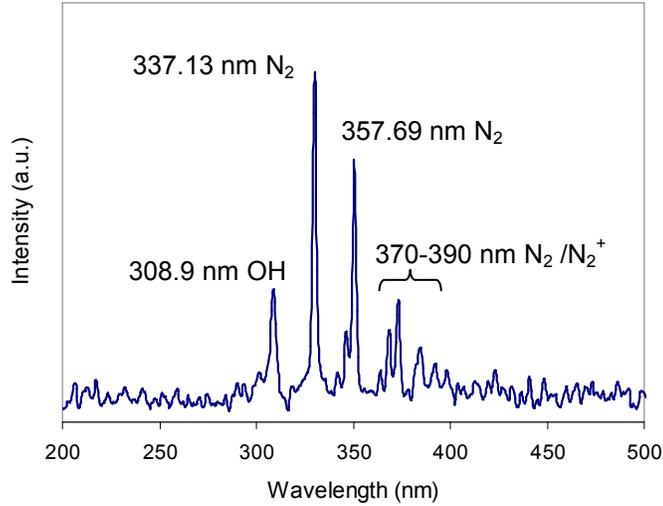

**FIG. 2.** Optical emission spectrum of a single plasma jet (air/He) operated downstream through the aerogel matrix. The main plasma species are identified.

The basic morphing capabilities of the CAPJ system have been tested by shaping the nozzle region with concave (Fig. 3a) and convex (Fig. 3b) geometries. To this end, a hard silica sphere of 10 mm in diameter has been used to shape conveniently the aerogel layer showing four apertures. The flexible dielectric resulted therefore in a static bent position towards the rear part (concave) or front part (convex) with a curvature radius of 5 mm. The four plasma jets were emerging from the convex nozzle with a significant divergence, even stronger than in the flat nozzle jet scenario (Fig. 1c). Here, the natural effect on jet bending due to the nozzle curvature is added to the earlier mentioned electrostatic repulsion between adjacent plasma plumes and the hydrodynamic effect due to gas expansion. In contrast, the CAP multi-jet system tended to merge at the nozzle axis in the concave configuration. In that case, only three jets from four apertures were active probably due to turbulence-generated instabilities. Indeed, one milestone consists of controlling the directionality of all the jets to enable uniform treatments on substrates. The lack of jet uniformity could be addressed by improving gas inlet distribution, and by using segmented electrodes biased at different voltages, being each voltage adapted to address jet directionality in any nozzle configuration.

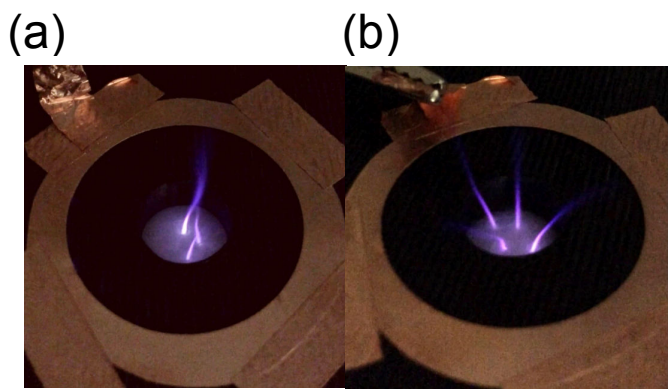

**FIG. 3.** Images of CAP multi-jets generated in **(a)** concave and **(b)** convex bending of the nozzle (5 mm in curvature radius).

In summary, the dielectric barrier performance of a flexible aerogel layer in a CAP multi-jet source has been demonstrated, thereby suggesting the construction of a morphing prototype based on this working principle. A study on the CAPJ directionality control (collimation) by exploring alternative electrode configurations and inlet distributions is envisaged to address non-uniformity issues in substrate treatments. In a further step, focused on unraveling the plasma chemistry associated with large-area CAPJ operation, we will learn the main reaction kinetics and chemical agents necessary to design smart plasma treatments for biomedical applications, like wound healing and treatment of surgical margins in cancer therapy. The scope of applications in plasma medicine will be definitely expanded by upgrading the conventional, "rigid" plasma sources into flexible sources adaptable to complex topographies.

**ACNOWLEDGEMENTS**

This work was supported by National Science Foundation through the award nr. 1919019. The authors acknowledge the assistance by Dr. Jiancun Rao from AIMLab at the Maryland NanoCenter.

**REFERENCES**

[1]  D. Yan, L. Lin, E. Gjika, C. Corbella, A. Malyavko, I.I. Beilis, J.H. Sherman, and M. Keidar, "Current Understanding of Mechanisms in Plasma Cancer Therapy and Recent Advances in Technology" in: Plasma Cancer Therapy, Ed. by M. Keidar, pp. 271-287 (Springer, Heidelberg, 2020)

[2]  G. Busco, E. Robert, N. Chettouh-Hammas, J.M. Pouvesle, and C. Grillon, "The emerging potential of cold atmospheric plasma in skin biology", Free Radic. Biol. Med. 161, 290-304 (2020).

[3]  G. Fridman, G. Friedman, A. Gutsol, A.B. Shekhter, V.N. Vasilets, and A. Fridman, "Applied plasma medicine", Plasma Process. Polym. 5, 503–533 (2008).

[4]  J. Golda, J. Held, B. Redeker, M. Konkowski, P. Beijer, A. Sobota, G. Kroesen, N.S.J. Braithwaite, S. Reuter, M.M. Turner, T. Gans, D. O'Connell, and V. Schulz-von der Gathen, "Concepts and characteristics of the 'COST reference microplasma jet'", J. Phys. D. Appl. Phys. 49, 084003 (2016).

[5]  K.D. Weltmann and T. Von Woedtke, "Plasma medicine—current state of research and medical application", Plasma Phys. Control. Fusion 59, 014031 (2017).

[6]  M. Laroussi, "Plasma Medicine: A Brief Introduction", Plasma 1, 47-60 (2018).

[7]  Q.Y. Nie, Z. Cao, C.S. Ren, D.Z. Wang, and M.G. Kong, "A two-dimensional cold atmospheric plasma jet array for uniform treatment of large-area surfaces for plasma medicine", New J. Phys. 11 115015 (2009).

[8] J. Kim, K.H. Choi, Y. Kim, B.J. Park, and G. Cho, "Wearable plasma pads for biomedical applications", Appl. Sci. 7, 1308 (2017).

[9] H. Jung, J.A. Seo, and S. Choi, "Wearable atmospheric pressure plasma fabrics produced by knitting flexible wire electrodes for the decontamination of chemical warfare agents", Sci. Rep. 7, 1–9 (2017).

[10] J. Xie, Q. Chen, P. Suresh, S. Roy, J.F. White, and A.D. Mazzeo, "Paper-based plasma sanitizers", Proc. Natl. Acad. Sci. 114, 5119–5124 (2017).

[11] P.P. Sun, E.M. Araud, C. Huang, Y. Shen, G.L. Monroy, S. Zhong, Z. Tong, S.A. Boppart, J.G. Eden, and T.H. Nguyen, "Disintegration of simulated drinking water biofilms with arrays of microchannel plasma jets", npj Biofilms Microbiomes 4, 24 (2018).

[12] T. Maho, X. Damany, S. Dozias, J. Pouvesle, and E. Robert, "Atmospheric pressure multijet plasma sources for cancer treatments", Clin. Plasma Med. 9, Supplement, 3–4 (2018).

[13] S. Bekeschus, P. Favia, E. Robert, and T. von Woedtke, "White paper on plasma for medicine and hygiene: Future in plasma health sciences", Plasma Process. Polym. 16, 1800033 (2019).

[14] Y. Lv, L. Nie, J. Duan, Z. Li, and X. Lu, "Cold atmospheric plasma jet array for transdermal drug delivery", Plasma Process. Polym. e2000180 (2020).

[15] J. Phalippou, T. Woignier, R. Sempere, and P. Dieudonne, "Highly porous aerogels of very low permeability", Mater. Sci. 20, 29-42 (2002).

[16] S. Ma, K. Kim, S. Lee, S. Moon, and Y. Hong, "Effects of a porous dielectric in atmospheric-pressure plasma jets submerged in water", Phys. Plasmas 25, 083519 (2018).

[17] L. Lin, Y. Lyu, B. Trink, J. Canady, and M. Keidar, "Cold atmospheric helium plasma jet in humid air environment ", J. Appl. Phys. 125, 153301 (2019).

[18] R. Xiong, Q. Xiong, A.Y. Nikiforov, P. Vanraes, and C. Leys, "Influence of helium mole fraction distribution on the properties of cold atmospheric pressure helium plasma jets", J. Appl. Phys. 112, 033305 (2012).

[19] M. Ghasemi, P. Olszewski, J.W. Bradley, and J.L. Walsh, "Interaction of multiple plasma plumes in an atmospheric pressure plasma jet array", J. Phys. D: Appl. Phys. 46, 052001 (2013).

[20] R.W.B. Pearse and A.G. Gaydon, The Identification of Molecular Spectra (Chapman & Hall Ltd., London, 1941).

[21] "NIST Atomic Spectra Database Lines Data", National Institute of Standards and Technology, see https://physics.nist.gov/PhysRefData/ASD/lines_form.html